\documentstyle[12pt]{article}

\newcommand {\m}{\mu}

\newcounter{eq}
\newcounter{sc}

\def\overleftrightarrow#1{\vbox{\ialign{##\crcr
 $\leftrightarrow$\crcr\noalign{\kern-1pt\nointerlineskip}
 $\hfil\displaystyle{#1}\hfil$\crcr}}}

\newlength{\minitwocolumn}
\begin{document}

\begin{flushright}
DPUR/TH/59\\
February, 2018\\
\end{flushright}
\vspace{20pt}

\pagestyle{empty}
\baselineskip15pt

\begin{center}
{\large\bf More on Weinberg's No Go Theorem in Quantum Gravity
\vskip 1mm }

\vspace{20mm}

Munehiro Nagahama\footnote{
           E-mail address:\ k178322@eve.u-ryukyu.ac.jp
                  }
and Ichiro Oda\footnote{
           E-mail address:\ ioda@phys.u-ryukyu.ac.jp
                  }

\vspace{10mm}
           Department of Physics, Faculty of Science, University of the 
           Ryukyus,\\
           Nishihara, Okinawa 903-0213, Japan\\

\end{center}


\vspace{10mm}
\begin{abstract}

We complement Weinberg's no go theorem on the cosmological constant problem in quantum gravity 
by generalizing it to the case of a scale-invariant theory. Our analysis makes use of the effective action and
the BRST symmetry in a manifestly covariant quantum gravity instead of the classical Lagrangian density and 
the $GL(4)$ symmetry in classical gravity. In this sense, our proof is very general since it does not depend on 
details of quantum gravity and holds true for general gravitational theories which are invariant under 
diffeomorphisms.  As an application of our theorem, we comment on an idea that in the asymptotic safety scenario 
the functional renormalization flow drives a cosmological constant to zero, solving the cosmological constant 
problem without reference to fine tuning of parameters.  Finally, we also comment on the possibility of extending 
the Weinberg theorem in quantum gravity to the case where the translational invariance is spontaneously broken.

\end{abstract}

\newpage
\pagestyle{plain}
\pagenumbering{arabic}


\rm
\section{Introduction}

The extremely tiny value of the cosmological constant at the present epoch, around $1 (meV)^4$, 
in the history of our universe poses a very serious problem to the community of both theoretical physics 
and cosmology \cite{Weinberg, Padilla}. This fact appears especially mysterious if it is assumed, as is probable, 
that the universe underwent several phase transitions which greatly change the cosmological constant.      
In the standard model of particle physics, spontaneous symmetry breaking naturally leads us to expect 
a cosmological constant of order $E^4$ where $E$ is the energy scale of symmetry breaking. This
energy scale ranges from $(100 MeV)^4$ for the QCD deconfinement phase transition to $(10^{18} GeV)^4$
for symmetry breaking at the Planck scale. Each phase transition yields a huge value of the cosmological 
constant, but the last phase transition must produce a very tiny cosmological constant to very high
accuracy. 

One can envision at least two scenarios of accounting for this tiny value of the cosmological constant.
One scenario is to appeal to some symmetry which reduces a large cosmological constant to the tiny one
or zero. The difficulty with the symmetry approach is that no symmetry is known at present which can do
such the job. The other is to utilize some dynamical mechanism which makes the cosmological
constant relax to the tiny value. A natural candidate realizing such a scenario is that some matter field
almost eats up the large cosmological constant and consequently a small fraction of it, which the matter field 
could not eat up, is left behind. It is this latter scenario that must confront and overcome the Weinberg's venerable 
no go theorem \cite{Weinberg, Padilla}.

It is well known that the existence of the cosmological constant makes it impossible for a flat Minkowski
metric to become a classical solution of the Einstein equation. This fact is upgraded to, what we call,
Weinberg's no go theorem, in an attempt of searching for a solution to the cosmological constant 
problem \cite{Weinberg}. The Weinberg's no go theorem in classical gravity can be stated as follows: 
General coordinate symmetries, or diffeomorphisms, which are in general violated by the presence of a fixed background metric, 
cannot be broken without any fine tuning of the effective cosmological constant such that the translational invariance, 
which is a subgroup of diffeomorphisms, is exactly preserved. In particular, any attempt relying on adjustment
mechanism, for which some matter field plays a role for erasing the cosmological constant, must be
confronted with and tried to evade the Weinberg theorem in order to provide a resolution
for the cosmological constant problem.

The Weinberg theorem also lays a cornerstone on recent developments of nonlocal approaches to the
cosmological constant problem where the operation of taking the space-time average of physical quantities 
plays a critical role, and as a result the effective cosmological constant is expressed in terms of
the space-time average of the trace of the energy-momentum tensor \cite{Linde}-\cite{Oda4}.\footnote{See
also related works \cite{Oda0}.}
 In the nonlocal approaches, one has to evaluate quantum effects of both matter and gravitational fields properly. 
It is therefore natural to extend the Weinberg's no go theorem, which is a purely classical statement 
based on field equations, to a quantum mechanical theorem. In this respect, note that the cosmological 
constant problem comes from a clash between particle physics and gravity in the semiclassical approximation 
where matter fields are quantized while gravity is treated as a classical theory. However, as seen 
in the nonlocal approaches to the cosmological constant problem, in order to describe the cosmological 
constant problem more accurately, it is necessary to take into consideration quantum effects 
from graviton loops, in other words, quantum gravity.    

In the previous work \cite{Oda5}, one of the authors has presented a quantum mechanical and nonperturbative
proof of the Weinberg theorem on the basis of the BRST transformation and the effective action, 
but its proof is restricted to the case where quantum field equations for matter and gravitational fields 
hold independently. In this article, we wish to generalize the proof to the case where the two field equations 
are dependent and related via a certain condition \cite{Weinberg}. The existence of this condition reflects 
the presence of scale symmetry in a theory, so the generalized Weinberg theorem can be applied to the situation 
where there is an exact scale symmetry within the framework of quantum gravity.  Since such a situation 
with an exact scale symmetry makes its appearance at fixed points in asymptotically safe theories of gravity 
\cite{ASG}\footnote{See also clear and concise review articles on asymptotic safety ideas \cite{Percacci}.}, 
it is expected that our theorem would provide some information for the approaches to the cosmological constant 
problem in asymptotically safe theories of gravity. Actually, there has recently appeared such a study 
where it was shown that the strong infrared effects of graviton's quantum fluctuations near the infrared 
fixed point solve the cosmological constant problem where the effective cosmological constant vanishes 
asymptotically in the limit of the infinite $\it{cosmon}$ field, $\chi  \rightarrow \infty$ \cite{Wetterich}. 
In this article, by using our Weinberg's no go theorem in quantum gravity, we point out that such a phenomenon 
might occur at not the infrared fixed point but the ultraviolet fixed point.
  
The rest of this paper is organised as follows: In the next section, we review Weinberg's no go theorem in
classical gravity. We see that the $GL(4)$ symmetry and the classical Lagrangian density are essential ingredients
for the proof of this theorem. In section 3, we present a proof of the Weinberg theorem in quantum 
gravity where the BRST symmetry and the effective action respectively play a similar role to the $GL(4)$ symmetry and 
the classical Lagrangian density.  Furthermore we comment on the relationship between our quantum theorem in
case of a scale-invariant theory and a model developed recently by Wetterich. Finally, we conclude in section 4.

\section{Weinberg's theorem in classical gravity}

In this section, let us review Weinberg's no go theorem in classical gravity \cite{Weinberg}. 
Arguments by Weinberg start with a Lagrangian density ${\cal L}(g_{\mu\nu}, \varphi_i)$ 
which consists of the metric tensor $g_{\mu\nu}$ and generic matter fields $\varphi_i$ 
where the subscript $i$ takes the values $i = 0, 1, 2, \cdots$ and labels different fields 
with suppressed tensor indices. Since we assume that the vacuum is translationally invariant, 
all fields must be constant fields on-shell, by which general coordinate symmetries, 
or diffeomorphisms, are reduced to a global $GL(4)$ symmetry \cite{Weinberg, Padilla}
\begin{eqnarray}
x^\mu \rightarrow x^{\prime\mu} = (M^{-1})^\mu \,_\nu x^\nu,
\label{GL(4)}
\end{eqnarray}
where $M^\mu \,_\nu$ is a constant $4 \times 4$ matrix satisfying $\det M \neq 0$. Then, we have
\begin{eqnarray}
g_{\mu\nu} &\rightarrow& g^\prime_{\mu\nu} = g_{\alpha\beta} M^\alpha \,_\mu M^\beta \,_\nu,
\nonumber\\
{\cal L}(g, \varphi_i) &\rightarrow& {\cal L}^\prime (g^\prime, \varphi^\prime_i) = \det M \cdot {\cal L}(g, \varphi_i). 
\label{GL(4)-2}
\end{eqnarray}
With $M^\mu \,_\nu = \delta^\mu_\nu + \delta M^\mu \,_\nu$ ($|\delta M^\mu \,_\nu| \ll 1$), the infinitesimal form
of the $GL(4)$ transformation is given by
\begin{eqnarray}
\delta g_{\mu\nu} = \delta M_{\mu\nu} +  \delta M_{\nu\mu}, \qquad
\delta {\cal L} = Tr \delta M \cdot {\cal L},
\label{GL(4)-3}
\end{eqnarray}
where $Tr \delta M \equiv g^{\mu\nu} \delta M_{\mu\nu}$.

Given constant fields on-shell, under the infinitesimal $GL(4)$ transformation, the Lagrangian density transforms as
\begin{eqnarray}
\delta {\cal L} = Tr \delta M \cdot {\cal L} = \frac{\partial {\cal L}}{\partial \varphi_i} \delta \varphi_i
+ \frac{\partial {\cal L}}{\partial g_{\mu\nu}} (\delta M_{\mu\nu} +  \delta M_{\nu\mu}).
\label{dL-CG1}
\end{eqnarray}
This relation is used to show that given the matter field equation $\frac{\partial {\cal L}}{\partial \varphi_i} = 0$,
the dependence of ${\cal L}$ on $g_{\mu\nu}$ is too simple to allow a solution of the gravitational field equation
$\frac{\partial {\cal L}}{\partial g_{\mu\nu}} = 0$.\footnote{When the matter fields are a scalar field $\varphi_i = \varphi$,
its $GL(4)$ variation is trivially zero, $\delta \varphi = 0$. Then, the first term on the RHS of Eq. (\ref{dL-CG1}) vanishes.
Even in this case, subsequent arguments are still valid.} To this end, let us consider two distinct cases separately; 
one case is that both of the field equations, those are, $\frac{\partial {\cal L}}{\partial \varphi_i} = 0$ and 
$\frac{\partial {\cal L}}{\partial g_{\mu\nu}} = 0$, hold independently whereas the other case is that they are not
independent and related to each other via a certain relation.

First, we will work with the former case where the two field equations are independent. Let us suppose that the matter field
equation is satisfied 
\begin{eqnarray}
\frac{\partial {\cal L}}{\partial \varphi_i} = 0.
\label{Matter-eq1}
\end{eqnarray}
Then, Eq. (\ref{dL-CG1}) is simply solved to be
\begin{eqnarray}
{\cal L} = \sqrt{-g} V(\varphi_i),
\label{Matter-sol1}
\end{eqnarray}
where $V(\varphi_i)$ is some function depending on only matter fields $\varphi_i$, which are a classical solution of 
Eq. (\ref{Matter-eq1}). As mentioned before, we see that the dependence of ${\cal L}$ on $g_{\mu\nu}$ is too simple to 
allow a solution of the gravitational field equation $\frac{\partial {\cal L}}{\partial g_{\mu\nu}} = 0$ unless $V(\varphi_i)$
is vanishing. Note that choosing $V(\varphi_i) = 0$ corresponds to fine tuning of the cosmological constant.

Next, let us turn our attention to the second case where the two field equations are related to each other through
a relation \cite{Weinberg}
\begin{eqnarray}
2 g_{\mu\nu} \frac{\partial {\cal L}}{\partial g_{\mu\nu}} = f_i (\varphi) \frac{\partial {\cal L}}{\partial \varphi_i},
\label{M-relation}
\end{eqnarray}
where $f_i (\varphi)$ is a certain function depending on $\varphi_i$. This relation can be rephrased as the existence of
a global symmetry in a theory
\begin{eqnarray}
\delta_\epsilon g_{\mu\nu} = 2 \epsilon  g_{\mu\nu}, \qquad  \delta_\epsilon \varphi_i = - \epsilon  f_i (\varphi).
\label{Scale-inv}
\end{eqnarray}
By using an appropriate redefinition of the fields $\varphi_i$, at least locally one can take \cite{Weinberg}
\begin{eqnarray}
\delta_\epsilon g_{\mu\nu} = 2 \epsilon  g_{\mu\nu}, \qquad  \delta_\epsilon \varphi = - \epsilon,
\qquad  \delta_\epsilon \varphi_a = 0,
\label{Scale-inv2}
\end{eqnarray}
where we have defined $\varphi = \varphi_0$ and $\varphi_a = \varphi_{i \neq 0}$.
This transformation coincides with "scale transformation" where $\varphi$ is a scalar field.\footnote{Precisely 
speaking, this transformation is not a conventional scale transformation since $\delta_\epsilon \varphi_a \neq 0$ 
for generic matter fields except the gauge field. For simplicity for the presentation, we will call Eq. (\ref{Scale-inv2}) 
scale transformation in this article.} Thus, it is natural to set its $GL(4)$ transformation to be zero, 
$\delta \varphi = 0$. Since we can show that
\begin{eqnarray}
\delta_\epsilon ( e^{2 \varphi} g_{\mu\nu} ) = 0,
\label{Scale-inv-metric}
\end{eqnarray}
the Lagrangian density, which is now scale-invariant, takes the form
\begin{eqnarray}
{\cal L} = {\cal L} ( e^{2 \varphi} g_{\mu\nu}, \varphi_a ) \equiv {\cal L} ( \hat g_{\mu\nu}, \varphi_a ),
\label{Scale-inv-Lag}
\end{eqnarray}
where we have introduced a scale-invariant metric $\hat g_{\mu\nu} = e^{2 \varphi} g_{\mu\nu}$.

We are willing to take the variation of ${\cal L}$ under the $GL(4)$ transformation
\begin{eqnarray}
\delta {\cal L} = \frac{\partial {\cal L}}{\partial \varphi_a} \delta \varphi_a
+ \frac{\partial {\cal L}}{\partial \hat g_{\mu\nu}} \delta \hat g_{\mu\nu},
\label{dL-CG2}
\end{eqnarray}
where $\delta \varphi = 0$ was used. Next, following the same line of reasoning as before, we impose
the matter field equation $\frac{\partial {\cal L}}{\partial \varphi_a} = 0$ without doing 
$\frac{\partial {\cal L}}{\partial \hat g_{\mu\nu}} = 0$, and as a result we obtain the relation
\begin{eqnarray}
\delta {\cal L} = Tr \delta \hat M \cdot {\cal L} = \frac{\partial {\cal L}}{\partial \hat g_{\mu\nu}} 
( \delta \hat M_{\mu\nu} +  \delta \hat M_{\nu\mu}),
\label{dL-CG3}
\end{eqnarray}
where we have defined $\hat M_{\mu\nu} = e^{2 \varphi} M_{\mu\nu}$,  $Tr \delta \hat M = \hat g^{\mu\nu} 
\delta \hat M_{\mu\nu}$ and $\hat g^{\mu\nu} = e^{-2 \varphi} g^{\mu\nu}$. The solution to Eq. (\ref{dL-CG3}) 
is given by
\begin{eqnarray}
{\cal L} = \sqrt{- \hat g} V(\varphi_a) = \sqrt{- g} e^{4 \varphi} V(\varphi_a).
\label{Matter-sol2}
\end{eqnarray}

The remaining field equation $\frac{\partial {\cal L}}{\partial \hat g_{\mu\nu}} = 0$ requires us to take
$V(\varphi_a) = 0$ or $e^{4 \varphi} \rightarrow 0$. As before, the former case corresponds to fine tuning
of the cosmological constant. On the other hand, the latter case is a new appearance, which deserves 
scrutiny. It is worthwhile to note that since the factor $e^{2 \varphi}$ always appears with $g_{\mu\nu}$,
all masses scale as  $e^\varphi$. The limit $e^{4 \varphi} \rightarrow 0$ then amounts to a massless and
scale-invariant theory. We are familiar with the fact that scale symmetry like (\ref{Scale-inv2}) is violated 
by conformal anomaly \cite{Weinberg},
by which the current conservation is violated.\footnote{The case of an exact scale invariance will be argued 
in the next section.} Consequently, we would obtain the effective Lagrangian density
\begin{eqnarray}
{\cal L}_{eff} =  {\cal L} + \sqrt{- g} \varphi \Theta^\mu _\mu,
\label{Effective-CL}
\end{eqnarray}
where $\Theta^\mu _\m$ represents the anomalous trace of the energy-momentum tensor. The existence of the 
term $\sqrt{- g} \varphi \Theta^\mu _\mu$ in the effective Lagrangian density means that we have no more 
scale symmetry at the quantum level, and this situation is then reduced to the former case where the two field equations
are independent and we need fine tuning of the cosmological constant.

\section{Weinberg's theorem in quantum gravity}

In this section, we wish to present a purely quantum mechanical proof of Weinberg's no go theorem.\footnote{The case
in the absence of scale symmetry has been already discussed in \cite{Oda5}.}  Before doing so, let us review briefly 
the manifestly covariant canonical formalism of quantum gravity \cite{Nakanishi1, Nakanishi-Ojima1}.

First, let us consider the conventional BRST transformation $\hat{\delta}_B$ for diffeomorphisms:
\begin{eqnarray}
\hat{\delta}_B g_{\mu\nu} &=& - c^\rho \partial_\rho g_{\mu\nu} - \partial_\mu c^\rho  g_{\rho\nu} - \partial_\nu c^\rho  g_{\rho\mu}
= - \nabla_\mu c_\nu - \nabla_\nu c_\mu, 
\nonumber\\
\hat{\delta}_B c^\mu &=& - c^\rho \partial_\rho c^\mu,   \quad
\hat{\delta}_B \bar c_\mu = i \hat{b}_\mu,  \quad   
\hat{\delta}_B \hat{b}_\mu = 0,  \quad
\hat{\delta}_B \varphi = - c^\rho \partial_\rho \varphi,   
\nonumber\\
\hat{\delta}_B A_\mu &=& - c^\rho \partial_\rho A_\mu - \partial_\mu c^\rho A_\rho,   \quad
\hat{\delta}_B x^\mu = 0,
\label{C-BRST}
\end{eqnarray}
where as the concrete example of matter fields we have given the BRST transformation for a real scalar field $\varphi$ 
and a vector field $A_\mu$.  This nilpotent BRST transformation $\hat{\delta}_B$ is obtained via the Lie derivative ${\cal{L}}_X$,
which is one-parameter family of diffeomorphisms generated by a vector field $X$, and commutes with $\partial_\mu$
since $\hat{\delta}_B$ does not change the values of coordinates.

The less familiar fact is that one can also define a different form of the nilpotent BRST transformation $\delta_B$ for diffeomorphisms:
\begin{eqnarray}
\delta_B g_{\mu\nu} &=& - \partial_\mu c^\rho  g_{\rho\nu} - \partial_\nu c^\rho  g_{\rho\mu},  \quad
\delta_B c^\mu = 0,  \quad   \delta_B \bar c_\mu = i b_\mu,  \quad   \delta_B b_\mu = 0,
\nonumber\\ 
\delta_B \varphi &=& 0,  \quad \delta_B A_\mu = - \partial_\mu c^\rho A_\rho,  \quad
\delta_B x^\mu = c^\mu.
\label{N-BRST}
\end{eqnarray}
It turns out that these two kinds of the BRST transformations are related by a relation
\begin{eqnarray}
\hat{\delta}_B \Phi = \delta_B \Phi - c^\rho \partial_\rho \Phi,
\label{BRST-relation}
\end{eqnarray}
where $\Phi \equiv \{ g_{\mu\nu}, \varphi, A_\mu, c^\mu,  \bar c_\mu, b_\mu, x^\mu \}$ and 
$\hat{b}_\mu \equiv b_\mu + i c^\rho \partial_\rho \bar c_\mu$ \cite{Nakanishi-Ojima1}.
Incidentally, the BRST transformation $\delta_B$ can be obtained from diffeomorphisms at the same
space-time point, so it does not commute with $\partial_\mu$; the commutator between them
is given by
\begin{eqnarray}
\delta_B ( \partial_\mu \Phi ) - \partial_\mu ( \delta_B \Phi) = - \partial_\mu c^\rho \partial_\rho \Phi.
\label{Non-comm}
\end{eqnarray}
Actually, one can verify Eq. (\ref{Non-comm}) in terms of Eq. (\ref{BRST-relation}) with the help of the
commutativity between $\hat \delta_B$ and $\partial_\mu$.

The two BRST transformations are mathematically equivalent so that one can use either at will.
However, for the present purpose, the BRST transformation $\delta_B$ is more convenient than the one
$\hat \delta_B$ since the Lagrangian density transforms as a density quantity under the BRST transformation 
$\delta_B$ as in the $GL(4)$ transformation in classical gravity. In this article, we therefore use the BRST 
transformation $\delta_B$ thoroughly.

Next, let us note that under the BRST transformation, $\sqrt{-g} d^4 x$ is the invariant measure, 
$\delta_B (\sqrt{-g} d^4 x) = 0$.  Then, the gauge-fixed and BRST-invariant action in the manifestly covariant
canonical formalism of quantum gravity is given by\footnote{It is straightforward to incorporate higher-derivative
terms such as $R^2$ and a scale-invariant coupling term, i.e. the nonminimal coupling term $\xi \varphi^2 R$ 
in this action.}
\begin{eqnarray}
S &=& \int d^4 x {\cal L}     \nonumber\\
&=& \int d^4 x \sqrt{- g} \left\{ \frac{1}{2} ( R - 2 \Lambda ) + \delta_B \left[ -i \bar c_\nu \frac{1}{\sqrt{- g}} 
\partial_\mu ( \sqrt{- g} g^{\mu\nu} ) \right] + \frac{1}{\sqrt{- g}} {\cal L}_m \right\}       \nonumber\\
&=& \int d^4 x \left[ \frac{1}{2} \sqrt{- g} ( R - 2 \Lambda ) + \partial_\mu ( \sqrt{- g} g^{\mu\nu} ) b_\nu 
-i \sqrt{- g} g^{\mu\nu} \partial_\mu \bar c_\rho \partial_\nu c^\rho + {\cal L}_m \right],
\label{BRST-action}
\end{eqnarray}
where we have put $\kappa^2 \equiv M_{Pl}^2 \equiv 8 \pi G_N = 1$ ($M_{Pl}$ is the reduced Planck mass and $G_N$
is Newton's constant), $R$ and $\Lambda$ are respectively the scalar curvature and a bare cosmological 
constant\footnote{We follow the conventions and notation by Misner et al. \cite{MTW}.}, and finally ${\cal L}_m$
is the Lagrange density for generic matter fields.  In this quantum action, as the gauge condition for diffeomorphisms, 
we have chosen the de Donder condition
\begin{eqnarray}
\partial_\mu ( \sqrt{- g} g^{\mu\nu} ) = 0. 
\label{de Donder1}
\end{eqnarray}
Owing to the identity $\partial_\nu ( \sqrt{- g} g^{\mu\nu} ) = - \sqrt{- g} g^{\nu\rho} \Gamma^\mu_{\nu\rho}$, 
Eq. (\ref{de Donder1}) can be rewritten as
\begin{eqnarray}
\Gamma^\mu_{\nu\rho}  g^{\nu\rho}  = 0, 
\label{de Donder2}
\end{eqnarray}
which is manifestly invariant under the general linear transformation $GL(4)$.
The action $S$ is not invariant under diffeomorphisms any longer, but it is still invariant under the BRST transformation 
and $GL(4)$ transformation. However, it is known that if the translational invariance is not spontaneously broken,
the $GL(4)$ is spontaneously broken in quantum gravity, by which one can prove that the graviton is exactly massless
owing to the Goldstone theorem \cite{Nakanishi-Ojima1, Nakanishi-Ojima2}.  Note that this proof of the masslessness of the
graviton is an exact proof without recourse to perturbation theory.  

Now we wish to present a quantum mechanical proof of the Weinberg theorem on the basis of the manifestly covariant 
canonical formalism of quantum gravity \cite{Oda5}.  To do that, it is worthwhile to recall what ingredients play a role in the proof of
the Weinberg theorem in classical gravity.  We find that they are the $GL(4)$ symmetry and the Lagrangian density as well as
the translational symmetry. The big issue is that the $GL(4)$ symmetry is spontaneously broken in quantum gravity so
we cannot use this symmetry any more. Instead, as an exact symmetry in quantum gravity, we have the BRST symmetry
stemming from diffeormorphisms, which is fully utilized in what follows.  Moreover, we have to ask ourselves if there is
an object having the whole quantum information in quantum gravity as in the Lagrangian density in classical gravity.
Indeed, as such an object we have the effective action $\Gamma [\varphi_i, g]$ which involves all information on radiative
corrections in addition to a classical action.   

In order to proceed in parallel with discussions on the $GL(4)$ symmetry, let us rewrite the BRST transformation of 
$g_{\mu\nu}$ as
\begin{eqnarray}
\delta_B g_{\mu\nu} = - \partial_\mu c^\rho  g_{\rho\nu} - \partial_\nu c^\rho  g_{\rho\mu} = \delta M_{\mu\nu} 
+ \delta M_{\nu\mu},
\label{g-BRST}
\end{eqnarray}
where we have defined
\begin{eqnarray}
M^\mu \, _\nu =  \delta^\mu _\nu + \lambda \delta M^\mu \, _\nu,
\label{delta-M}
\end{eqnarray}
with $\lambda$ being a Grassmann-odd parameter, which will be omitted henceforth since it is irrelevant to
later arguments. At this stage, let us consider an integrand of the effective action, $\tilde \Gamma$ which is defined as
\begin{eqnarray}
\Gamma = \int d^4 x \ \tilde \Gamma,
\label{tilde EA}
\end{eqnarray}
where the effective action $\Gamma$ is assumed to be invariant under the BRST transformation (\ref{N-BRST}).  
Since $\tilde \Gamma$ transforms as a density quantity under the BRST transformation, we have
\begin{eqnarray}
\tilde \Gamma \rightarrow \tilde \Gamma^\prime = (\det M) \tilde \Gamma,
\label{BRST tilde EA}
\end{eqnarray}
and its infinitesimal form reads
\begin{eqnarray}
\delta_B \tilde \Gamma = \tilde \Gamma^\prime - \tilde \Gamma \approx (Tr \delta M) \tilde \Gamma
= - (\partial_\rho c^\rho) \tilde \Gamma.
\label{inf-BRST tilde EA}
\end{eqnarray}
In fact, using Eq. (\ref{inf-BRST tilde EA}) and $\delta_B \sqrt{-g} = - \sqrt{-g} \partial_\rho c^\rho$, 
$\frac{1}{\sqrt{-g}} \tilde \Gamma$ is found to be invariant under the BRST transformation, thereby showing that 
the effective action $\Gamma$ in Eq. (\ref{tilde EA}) is BRST-invariant as the integration measure $\sqrt{-g} d^4 x$
is also BRST-invariant.

Now let us assume that the translational invariance is not spontaneously broken, which indicates that 
\begin{eqnarray}
\langle 0| g_{\mu\nu} | 0 \rangle = \eta_{\mu\nu}, \qquad \langle 0| \varphi_i | 0 \rangle = \varphi_i^{(0)},
\label{VEV of g}
\end{eqnarray}
where $| 0 \rangle$ denotes the true vacuum state, and $\eta_{\mu\nu}$ and $\varphi_i^{(0)}$ are respectively
a flat Minkowski metric and constant modes, both of which are independent of the space-time coordinates. 
The vacuum expectation value of the remaining fields, those are, the auxiliary field $b_\mu$, the FP ghost $c^\mu$ and 
the FP antighost $\bar c_\mu$, is taken to be zero owing to the Lorentz invariance and the conservation of ghost
number.  Next, let us take the BRST variation of the effective action
\begin{eqnarray}
\delta_B \tilde \Gamma = \frac{\partial \tilde \Gamma}{\partial \varphi_i} \delta_B \varphi_i 
+ \frac{\partial \tilde \Gamma}{\partial g_{\mu\nu}} \delta_B g_{\mu\nu} 
+ \frac{\partial \tilde \Gamma}{\partial c^\mu} \delta_B c^\mu
+ \frac{\partial \tilde \Gamma}{\partial \bar c_\mu} \delta_B \bar c_\mu
+ \frac{\partial \tilde \Gamma}{\partial b_\mu} \delta_B b_\mu. 
\label{Relation}
\end{eqnarray}

Meanwhile, for the translational invariant fields, the quantum field equations, or the generalized Euler-Lagrange equations, 
which involve all quantum effects, are given by
\begin{eqnarray}
\frac{\partial \tilde \Gamma}{\partial \varphi_i}
= \frac{\partial \tilde \Gamma}{\partial g_{\mu\nu}} = \frac{\partial \tilde \Gamma}{\partial c^\mu}
= \frac{\partial \tilde \Gamma}{\partial \bar c_\mu} = \frac{\partial \tilde \Gamma}{\partial b_\mu} 
= 0.
\label{q-eq}
\end{eqnarray}
By imposing $\frac{\partial \tilde \Gamma}{\partial c^\mu} = \frac{\partial \tilde \Gamma}{\partial \bar c_\mu} 
= \frac{\partial \tilde \Gamma}{\partial b_\mu} = 0$ on Eq. (\ref{Relation}) (indeed, $\frac{\partial \tilde \Gamma}{\partial \bar c_\mu}
= 0$ is sufficient since $\delta_B c^\mu = \delta_B b_\mu = 0$), we obtain 
\begin{eqnarray}
\delta_B \tilde \Gamma = \frac{\partial \tilde \Gamma}{\partial \varphi_i} \delta_B \varphi_i 
+ \frac{\partial \tilde \Gamma}{\partial g_{\mu\nu}} \delta_B g_{\mu\nu}. 
\label{Relation2}
\end{eqnarray}
As in classical gravity, there are two possibilities, one of which is that quantum field equations 
$\frac{\partial \tilde \Gamma}{\partial \varphi_i} = 0$ and $\frac{\partial \tilde \Gamma}{\partial g_{\mu\nu}} = 0$ 
are independent while the other is that they are not independent and related by a condition.

As before, let us treat with the two cases separately.  When the two quantum field equations are independent,
we first set $\frac{\partial \tilde \Gamma}{\partial \varphi_i} = 0$. Then, Eqs. (\ref{N-BRST}), (\ref{inf-BRST tilde EA}) 
and (\ref{Relation2}) give rise to a relation
\begin{eqnarray}
\delta_B \tilde \Gamma = \frac{\partial \tilde \Gamma}{\partial g_{\mu\nu}} \delta_B g_{\mu\nu}
= \frac{\partial \tilde \Gamma}{\partial g_{\mu\nu}} (- 2 \partial_\mu c^\rho g_{\rho\nu})
= - ( \partial_\rho c^\rho ) \tilde \Gamma,
\label{New Relation}
\end{eqnarray}
which provides us with the equation
\begin{eqnarray}
\frac{\partial \tilde \Gamma}{\partial g_{\mu\nu}} - \frac{1}{2} g^{\mu\nu} \tilde \Gamma = 0.
\label{Final eq}
\end{eqnarray}
This equation can be easily solved to be
\begin{eqnarray}
\tilde \Gamma = \sqrt{-g} V(\varphi_i),
\label{Final ans}
\end{eqnarray}
where $V(\varphi_i)$ is some function of only $\varphi_i$.  Note that $V$ is nothing but the effective potential since the vacuum
expectation values of fields do not depend on the space-time coordinates.  
Finally, requiring $\frac{\partial \tilde \Gamma}{\partial g_{\mu\nu}} = 0$ leads to
\begin{eqnarray}
V(\varphi_i) = 0,
\label{V=0}
\end{eqnarray}
which corresponds to fine tuning of the cosmological constant at the level of quantum gravity. In this way, we have succeeded in proving
a quantum mechanical generalization of the Weinberg's no go theorem when the quantum field equations for matter and gravitational 
fields are independent \cite{Oda5}.

Next, we will move on to the case where $\frac{\partial \tilde \Gamma}{\partial g_{\mu\nu}} = 0$ is related to 
$\frac{\partial \tilde \Gamma}{\partial \varphi_i} = 0$ via a relation
\begin{eqnarray}
2 g_{\mu\nu} \frac{\partial \tilde \Gamma}{\partial g_{\mu\nu}} = f_i (\varphi) \frac{\partial \tilde \Gamma}{\partial \varphi_i}.
\label{M-relation2}
\end{eqnarray}
In this case, we can also proceed in a similar way to the case of classical gravity. Let us first consider the BRST variation 
(of the integrand) of the effective action
\begin{eqnarray}
\delta_B \tilde \Gamma = \frac{\partial \tilde \Gamma}{\partial \varphi_i} \delta_B \varphi_i 
+ \frac{\partial \tilde \Gamma}{\partial g_{\mu\nu}} \delta_B g_{\mu\nu}. 
\label{Relation3}
\end{eqnarray}
Owing to the scale symmetry (\ref{Scale-inv2}) coming from the relation (\ref{M-relation2}), 
the effective action, which should be scale-invariant, must be a function of the scale-invariant metric $\hat g_{\mu\nu}
= e^{2 \varphi} g_{\mu\nu}$ and matter fields $\varphi_a$:
\begin{eqnarray}
\tilde \Gamma = \tilde \Gamma ( e^{2 \varphi} g_{\mu\nu}, \varphi_a ) = \tilde \Gamma ( \hat g_{\mu\nu}, \varphi_a ).
\label{Scale-inv-eff-action}
\end{eqnarray}

The key observation here is that as seen in Eq. (\ref{N-BRST}), the scalar field $\varphi$ is invariant under the BRST 
transformation, $\delta_B \varphi = 0$, which should be contrasted to the case of the different form of the BRST 
transformation, $\hat  \delta_B \varphi = - c^\rho \partial_\rho \varphi$ in Eq. (\ref{C-BRST}).  This fact is another 
advantage of the BRST transformation $\delta_B$ over the one $\hat  \delta_B$. Thus, the BRST transformation of 
the effective action now takes the form
\begin{eqnarray}
\delta_B \tilde \Gamma = \frac{\partial \tilde \Gamma}{\partial \varphi_a} \delta_B \varphi_a
+ \frac{\partial \tilde \Gamma}{\partial \hat g_{\mu\nu}} \delta_B \hat g_{\mu\nu}.
\label{dL-QG}
\end{eqnarray}
Assuming $\frac{\partial \tilde \Gamma}{\partial \varphi_a} = 0$ without imposing 
$\frac{\partial \tilde \Gamma}{\partial \hat g_{\mu\nu}} = 0$, Eq. (\ref{dL-QG}) can be cast to
\begin{eqnarray}
\delta_B \tilde \Gamma = \frac{\partial \tilde \Gamma}{\partial \hat g_{\mu\nu}} \delta_B \hat g_{\mu\nu}
= \frac{\partial \tilde \Gamma}{\partial \hat g_{\mu\nu}} ( - 2 \partial_\mu c^\rho \hat g_{\rho\nu} )
= - ( \partial_\rho c^\rho ) \tilde \Gamma,
\label{dL-QG2}
\end{eqnarray}
where we have used Eq. (\ref{inf-BRST tilde EA}) at the last step. This relation gives rise to the equation
for the effective action $\tilde \Gamma$:
\begin{eqnarray}
\frac{\partial \tilde \Gamma}{\partial \hat g_{\mu\nu}} - \frac{1}{2} \hat g^{\mu\nu} \tilde \Gamma = 0,
\label{QG-eq}
\end{eqnarray}
which is solved to be 
\begin{eqnarray}
\tilde \Gamma = \sqrt{- \hat g} V(\varphi_a) = \sqrt{- g} e^{4 \varphi} V(\varphi_a).
\label{QMatter-sol}
\end{eqnarray}
The remaining quantum field equation $\frac{\partial \tilde \Gamma}{\partial g_{\mu\nu}} = 0$ forces us to select
$V(\varphi_a) = 0$ or $e^{4 \varphi} \rightarrow 0$, which is the same condition as in classical gravity. Indeed, the former case 
$V(\varphi_a) = 0$ precisely corresponds to fine tuning of the cosmological constant in quantum gravity as before. 

However, the physical interpretation of the latter choice $e^{4 \varphi} \rightarrow 0$ at hand is different from that in 
classical gravity, so let us investigate this case more closely. Since we deal with the effective action involving all radiative 
corrections (in addition to the classical action), the limit  $e^{4 \varphi} \rightarrow 0$  in quantum gravity results in
an exactly scale-invariant theory where there is no conformal anomaly and the $\beta$ function is identically vanishing 
as in the well-known $N=4$ super Yang-Mills theory \cite{Mandelstam}. Indeed, as mentioned before, owing to scale symmetry, 
the metric tensor $g_{\mu\nu}$ always comes with the factor $e^{2 \varphi}$, so all masses scale as $e^\varphi$.
This fact is, for instance, verified by looking at the scale-invariant kinetic term of a scalar field as follows:
\begin{eqnarray}
\frac{1}{2} m^2 \sqrt{- \hat g} \hat g^{\mu\nu} \partial_\mu \varphi \partial_\nu \varphi 
= \frac{1}{2} (m e^\varphi)^2 \sqrt{- g} g^{\mu\nu} \partial_\mu \varphi \partial_\nu \varphi, 
\label{Scalar-kinetic}
\end{eqnarray}
where the effective mass is found to be $m_{eff} = m e^\varphi$ as expected. Thus, it is true that the limit  
$e^{4 \varphi} \rightarrow 0$ corresponds to a massless, scale-invariant fixed point. The next important work is to check 
whether this fixed point is the infrared fixed point or the ultraviolet one. To understand it, let us focus on 
the scale-invariant line element
\begin{eqnarray}
d \hat s^2 \equiv \hat g_{\mu\nu} d x^\mu d x^\nu = e^{2 \varphi} g_{\mu\nu} d x^\mu d x^\nu
\equiv e^{2 \varphi} d s^2, 
\label{Line}
\end{eqnarray}
from which it turns out that the limit  $e^{4 \varphi} \rightarrow 0$ is the ultraviolet limit since
the scale-invariant line element $d \hat s$ shrinks to zero in this limit for a fixed line element $d s$.  
To put differently, the limit  $e^{4 \varphi} \rightarrow 0$ implies that in this limit we are supposed to approach 
the ultraviolet (UV-)fixed point. Accordingly, at the UV-fixed point, we are free from the issue of fine tunig 
of the cosmological constant problem. This result is quite reasonable since it gives us the following physically
plausible picture; at high energies, an exact scale symmetry is in general restored and consequently it forbids the 
effective cosmological constant to emerge in a theory. However, the real issue in this case is that it seems 
to be difficult to construct a realistic cosmological model at low energies on the basis of such an exactly scale-invariant theory 
in the ultraviolet regime since the large cosmological constant is expected to reappear at low energies owing to 
the lack of an exact scale symmetry.  

Incidentally, the infrared (IR-)fixed point has been recently used to make the effective cosmological constant be asymptotically 
vanishing by Wetterich where a scalar field called the $\it{cosmon}$ field $\chi$ plays an important role \cite{Wetterich}.  
In order to extract a relation between the cosmon field $\chi$ and our scalar filed $\varphi$, let us consider the nonminimal
coupling term by beginning with the scale-invariant Einstein-Hilbert action
\begin{eqnarray}
{\cal L}_{EH} \equiv \sqrt{- \hat g} \hat R = \sqrt{- g} e^{2 \varphi} R \equiv \sqrt{- g} \chi^2 R,
\label{EH}
\end{eqnarray}
where we have used the formulae for scale transformation $\hat g_{\mu\nu} = \Omega^2 g_{\mu\nu}$, those are, 
$\hat R = \Omega^{-2} R, \sqrt{- \hat g} = \Omega^4  \sqrt{- g}$ for a constant scale factor $\Omega$. 
From Eq. (\ref{EH}), we can read out the relation between the two scalar fields
\begin{eqnarray}
\chi = e^\varphi.
\label{chi}
\end{eqnarray}
Thus, the ultraviolet limit $e^\varphi \rightarrow 0$ turns out to correspond to $\chi \rightarrow 0$  while the infrared limit 
$e^\varphi \rightarrow \infty$ does $\chi \rightarrow \infty$. 
The present proof of the Weinberg's no go theorem in quantum gravity casts some doubt on the Wetterich's idea 
such that the IR-fixed point at $\chi \rightarrow \infty$ might provide a useful method in the cosmological constant problem since  
according to our theorem it is the UV-fixed point that erases the cosmological constant without fine tuning.  
Even if his idea were true by violating some requirement of our no go theorem tacitly, it is not clear whether one could 
construct a realistic model describing our universe from an exactly scale-invariant theory in the infrared regime since our
universe is never scale-invariant and most of elementary particles are massive except a few gauge particles at low energies.

\section{Discussion}

In this article, we have complemented Weinberg's no go theorem on the cosmological constant problem in quantum gravity 
by including the case of a scale-invariant theory. Our proof is very general in the sense that we use only the BRST symmetry
and the BRST-invariant effective action and the detail of quantum gravity is almost irrelevant.

The cosmological constant problem stems from a clash between particle physics which sources the vacuum energy density 
through radiative corrections and gravity which responds to the vacuum energy density classically. 
It is therefore not obvious ${\it a \ priori}$ whether quantum gravity plays a role in the cosmological constant
problem. In addition to it, a widespread belief among us is that quantum gravity effects become dominant only at the
Planck scale region where the semiclassical approach breaks down.  However, recent studies suggest that quantum gravity
might give us non-negligible contributions to the cosmological constant via some enhancement mechanism of quantum gravity
effects. Under such a situation, the present theorem would provide a criterion on model building for the cosmological constant
problem based on quantum gravity.

As an application of our theorem, we have referred to a recent work that strong ${\it infrared}$ quantum gravity effects give rise to
an asymptotically vanishing cosmological constant. Our theorem implies that ${\it ultraviolet}$ quantum gravity effects,
together with an exact scale symmetry, lead to an asymptotically vanishing cosmological constant without fine tuning, 
but it seems that the real issue lies in how to build a realistic cosmological model since our world does not possess 
such scale symmetry at least in the low energy region. 

As a final comment, we would like to mention the possibility of extending a recent work \cite{Niedermann}, in which original 
Weinberg's arguments in classical gravity have been generalized to the case with the broken translation and non-local kinetic operator, 
to quantum gravity. In this case, since the translational symmetry is spontaneously broken, the vacuum expectation value 
$\langle 0| g_{\mu\nu} | 0 \rangle$ is not a constant tensor, implying a curved background geometry.  Then,
the $GL(4)$ symmetry is completely broken and as a result not only the covariant conservation law of the energy-mometum tensor 
but also the Lorentz symmetry are broken.\footnote{In case that the translational symmetry is not spontaneously 
broken, one can show in a nonperturbative manner that the $GL(4)$ symmetry is spontaneously broken to the Lorentz group,
thereby making it possible for the graviton to be exactly massless owing to the Goldstone theorem \cite{Nakanishi-Ojima2}.}
In such a situation, the scenario of creation of the universe from nothing would become realistic. From this perspective, 
it seems to be difficult to establish Weinberg's no go theorem in quantum gravity for the case where the translational 
symmetry is spontaneously broken.

\begin{flushleft}
{\bf Acknowledgements}
\end{flushleft}
The work of I. O.  was supported by JSPS KAKENHI Grant Number 16K05327.


\end{document}